\documentclass[a4paper,fleqn]{cas-dc}
\usepackage{url}
\usepackage{graphicx}
\usepackage{caption}
\usepackage{subcaption}
\usepackage{lipsum}
\usepackage{natbib}
\usepackage{amsmath}
\def\tsc#1{\csdef{#1}{\textsc{\lowercase{#1}}\xspace}}
\tsc{WGM}
\tsc{QE}
\tsc{EP}
\tsc{PMS}
\tsc{BEC}
\tsc{DE}
\newcommand\arcsec{\mbox{$^{\prime\prime}$}}%
\newcommand\fp{\mbox{$.\!\!^{\scriptscriptstyle\mathrm p}$}}%
\def\farcm{%
 \mbox{.\kern -0.7ex\raisebox{.9ex}{\scriptsize$\prime$}}%
}%
\def\farcs{%
 \mbox{%
  \kern  0.13ex.%
  \kern -0.95ex\arcsec%
  \kern -0.1ex%
 }%
}%
\makeatletter
\newcommand\ion[2]{#1$\;$\protect \textsmaller{\rmfamily\@Roman{#2}}}%
\newcommand\smion[2]{#1$\;$\protect \footnotesize{\rmfamily\@Roman{#2}}\small}%
\newcommand\scion[2]{#1$\;$\protect \tiny{\rmfamily\@Roman{#2}}\scriptsize}%
\makeatother

\begin{document}
\let\WriteBookmarks\relax
\def\floatpagepagefraction{1}
\def\textpagefraction{.001}

\shorttitle{Reconstructing 3-D Jet Geometry}

\shortauthors{Sawant, Kosak, Li, Avachat et al.}

\title [mode = title]{{\sl JetCurry} I. Reconstructing Three-Dimensional Jet Geometry from Two-Dimensional Images}                      

\author[1]{Sailee M. Sawant}[orcid = 0000-0002-7987-0310]
\ead{ssawant2011@my.fit.edu}
\fnmark[1]

\author[1, 3]{Katie Kosak}[orcid = ]
\fnmark[1]
\author[1, 4]{Kunyang Li}[orcid = 0000-0002-0867-8946]
\fnmark[1]
\author[1, 5]{Sayali S. Avachat}[orcid = 0000-0002-5550-5693]
\fnmark[1]
\author[1]{Eric S. Perlman}[orcid=0000-0002-3099-1664]
\author[2]{Debasis Mitra}[orcid=0000-0002-4351-1252]

\affiliation[1]{organization = {Department of Aerospace, Physics and Space Sciences, Florida Institute of Technology},
    addressline={150 W. University Blvd.}, 
    city={Melbourne},
    state={FL},
    postcode={32901}, 
    country={USA}}
\affiliation[2]{organization = {Department of Computer Engineering and Sciences, Florida Institute of Technology},
    addressline={150 W. University Blvd.}, 
    city={Melbourne}, 
    state={FL},
    postcode={32901}, 
    country={USA}}
\affiliation[3]{organization = {Physics Department, University of Warwick},
    addressline = {Coventry},
    postcode={CV4 7AL}, 
    country={UK}}
\affiliation[4]{organization = {Center for Relativistic Astrophysics, School of Physics, Georgia Institute of Technology},
    addressline={837 State Street}, 
    city={Atlanta},
    state={GA},
    postcode={30332}, 
    country={USA}}
    
\affiliation[5]{organization = {Inter-University Centre for Astronomy and Astrophysics},
    city={Pune},
    state={Maharashtra},
    postcode={411007}, 
    country={India}}


\fntext[fn1]{These authors contributed equally.}

\begin{abstract}
We present a three-dimensional (3-D) visualization of jet geometry using numerical methods based on a Markov Chain Monte Carlo (MCMC) and limited memory Broyden--Fletcher--Goldfarb--Shanno (BFGS) optimized algorithm. Our aim is to visualize the 3-D geometry of an active galactic nucleus (AGN) jet using observations, which are inherently two-dimensional (2-D) images. Many AGN jets display complex structures that include hotspots and bends. The structure of these bends in the jet's frame may appear quite different than what we see in the sky frame, where it is transformed by our particular viewing geometry. The knowledge of the intrinsic structure will be helpful in understanding the appearance of the magnetic field and hence emission and particle acceleration processes over the length of the jet. We present the {\sl JetCurry} algorithm to visualize the jet's 3-D geometry from its 2-D image. We discuss the underlying geometrical framework and outline the method used to decompose the 2-D image.  We report the results of our 3-D visualization of the jet of M87, using the test case of the knot D region. Our 3-D visualization is broadly consistent with the expected double helical magnetic field structure of knot D region of the jet. We also discuss the next steps in the development of the {\sl JetCurry} algorithm.
\end{abstract}
\maketitle
\section{Introduction}
\indent Relativistic jets transport mass and energy from sub-parsec central regions to Mpc-scale lobes, with a kinetic power comparable to that of their host galaxies and active galactic nuclei (AGNs). This profoundly influences the evolution of the hosts, nearby galaxies, and the surrounding interstellar and intracluster medium \citep{2012A&A...545L..11S, 2012ARA&A..50..455F}. The generation of such flows is tied to the process of accretion onto (likely) rotating black holes, where the magneto-rotational instability can couple the black hole's spin and magnetic field to the disk or flow to produce high-latitude outflows at speeds close to the speed of light \citep{2001Sci...291...84M}. While these jets have a dominant direction of motion (i.e., outward from the black hole), they often have bends and features (both stationary and moving) that are either perpendicular or aligned relative to the jet at some angle. Deciphering the true nature of these features and their geometry, relation, and dynamical meaning within the flow is a difficult problem, as any astronomical images we acquire are of necessity two-dimensional (2-D) views of three-dimensional (3-D) objects.

\begin{figure}[t!]
\centering
\includegraphics[width=\linewidth]{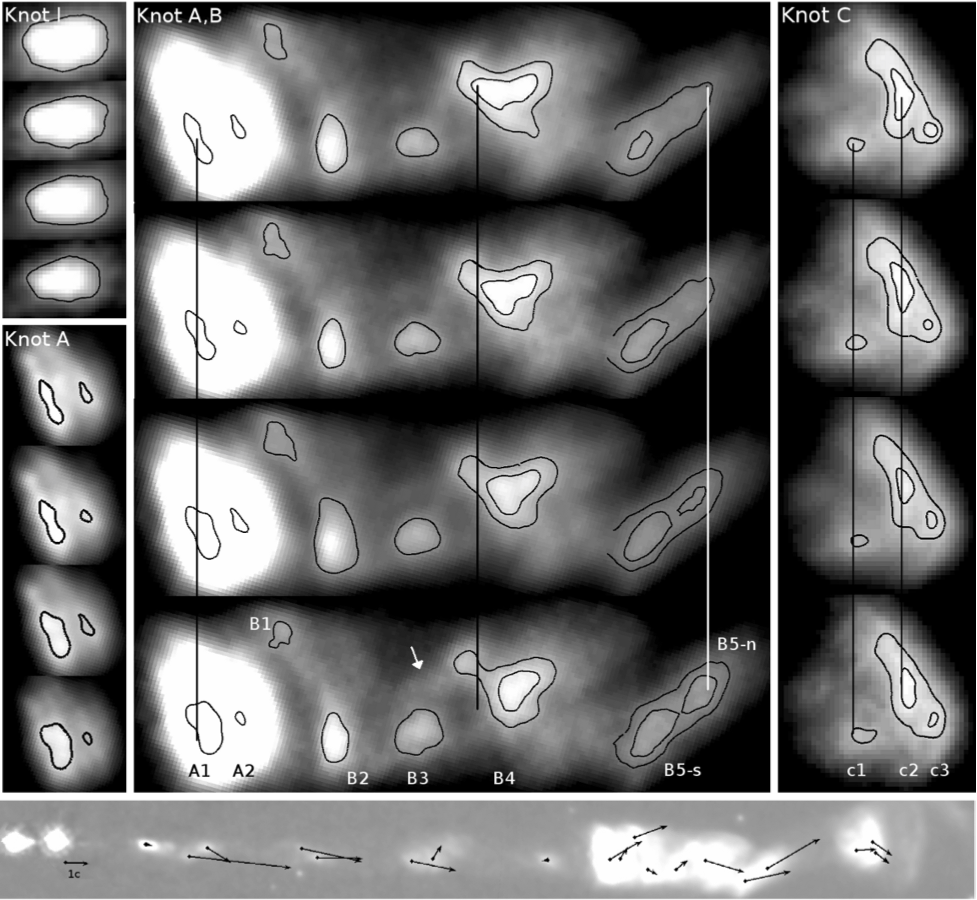}
\caption{Superluminal motion of the sub-components in several of the knot regions (I, A, B, and C) of the M87 jet, spanning over 13.25 years of monitoring between 1995-2008, with the {\it Hubble Space Telescope.} The westward direction lies 20.5$^\circ$ below the horizontal.  The bottom panel depicts the velocities as vectors from their positions in the jet. The length of the vectors is proportional to the apparent speed.  Reproduction of Figure 3 from \citet{Meyer2013}. Courtesy--Eileen Meyer.}
\label{fig:vectors-meyer}
\end{figure}

\indent The problem of reconstructing 3-D  information from 2-D images is common to many fields, but it is particularly critical in astronomy.  In most other cases, e.g., medical imaging, one may take images of a source from multiple viewpoints to aid reconstruction.  However, this is not possible in astronomy, so we must rely on other methods.  For instance, \citet{2011ITVCG..17..454S}, \citet{2012arXiv1204.6132W}, \citet{wenger2013fast}, \citet{cormier13}, \cite{SabatiniEtAl2018}, and \citet{LagattutaEtAl2019} used symmetries inherent in, respectively, planetary nebulae and galaxies, plus 2-D images, to infer and reconstruct 3-D visualizations of these objects.  This field is, in fact, rapidly growing in astronomy, as can be seen by the vast number of subjects explored on the 3DAstrophysics blog\footnote[2]{https://3dastrophysics.wordpress.com/}. \\
\indent In astrophysical jets, the problem is rather different. Unlike in galaxies or planetary nebulae, we cannot make assumptions such as spherical, elliptical or disk symmetry, or rotation.  However,  we can assume a dominant direction of propagation.  As an example of the typical knotted structure of AGN jets, we show in Figure~\ref{fig:vectors-meyer}, a broad view of the M87 jet, one of the nearest of the class at 16.7 Mpc distance, taken from \citet{Meyer2013}.  In every single image, the M87 jet shows an amazing complexity of features, including knots, helical undulations, shocks, and a variety of other morphological structures, many of which are oriented at some odd angles with respect to the overall jet direction. As shown by \citet{Meyer2013}, some of the features in the M87 jet seem to move with apparent velocities up to about 6$c$ within the inner $12\farcs0$ of the jet, with a general decline in apparent speed with increasing distance from the nucleus. However, there are some nearly stationary components that are largely located near the upstream ends of knots. Additionally, the polarimetric imaging of \citet{2016ApJ...832....3A} shows apparent helical winding structures to the inferred  magnetic field vectors in several knots. These features are clues to complex jet dynamics, but are difficult to interpret properly.

\indent In this paper, we describe a geometrically based code that attempts to reconstruct the 3-D structure of jets, starting from 2-D images.
This is an evolving project that in later stages will attempt to use kinematic information as well as incorporate special relativistic corrections so that foreshortening, Doppler boosting, and superluminal motion can be included. 
Our goal is to provide a firmer geometrical grounding to these modeling efforts by allowing reconstruction of a jet's structure 
in 3-D.

\section{Geometrical Framework} \label{sec:model}
To visualize the geometry of the jet in 3-D, the key parameters to consider are the distance between any two features and the apparent angle between them with respect to the direction of the jet’s axis. The distance between the core of the jet and the location of interest on the jet is defined as $s$, and the angle of the location to the core is $\eta$. Another parameter is the line of sight (LOS). 

We assumed that the 2-D projection of the jet’s axis lies along the $x$-axis and measured the angles with respect to the positive $x$ direction. We considered $s$ and $\eta$ as our known parameters, which can be directly measured from the images, i.e., in the sky frame. A third parameter, assumed to be known (albeit from other information such as a  $\beta$ vs. $\theta$ plot based on the observed superluminal motion, where $\beta$ is the space velocity and $\theta$ is the viewing angle with respect to the LOS) is the angle the jet's propagation axis makes with respect to the LOS.

Following \citet{1993ApJ...411...89C}, Figure \ref{fig:geometry_p} shows the relevant geometry for a single bend within a jet, and specifically how the 2-D sky frame can be related to the jet's frame, which is inherently 3-D. All primed points represent the observed, sky-frame projection we see, with the components lying at A$'$ and B$'$ in that frame but at points A and B in the jet's frame. The point D$'$ is the projection of B$'$ on the $+x$ axis, and the point D is projection of B on ($x, z$) plane. Angle $\eta$ is the angle between segment A$'$B$'$ and the $+x$ axis. From point B draw a line BC perpendicular to the jet axis (OC) and set the $\angle$CAB as $\xi$. Point C is then where line BC crosses line OC perpendicularly, so $\angle$BCA is 90$^{\circ}$. Segment BC makes an angle $\phi$ with the ($x, z$) plane (i.e., $\angle$BCD = $\phi$). This way, $\Delta$ABC is raised off the ($x, z$) plane through angle $\phi$, while the segment AC still lies in the $(x,z)$ plane. Segment AC makes an angle $\theta$ with the LOS, which is assumed to be along the $+z$ axis.  The distance between points A and B in the jet's frame is $d$, while $s$ is the projection of $d$ on the ($x, y$) plane, i.e., the distance between A$'$ and B$'$.  Angle $\alpha$ is the apex angle of $\Delta$BAD, and $\beta$ is the angle between triangles BAD and FAE. Finally, $\Delta$AGH is the projection of $\Delta$ABD on the ($y, z$) plane.

To simplify the algorithm both computationally and physically, for now we assume that the jet is non-relativistic. The LOS effects in addition to various relativistic effects can enhance the intensity and shift the frequency observed, as well as change the comparison between geometry in the jet and observer frames \citep{Bottcher_Chapter_2}. These effects will be included in the next version of {\sl JetCurry}.

\subsection{Non-linear Parametrized Equations} \label{sec:non-lin-eq}
We used a set of non-linear parametrized equations containing the angles and distances described above. Assuming the non-relativistic jet flow and using the geometry in Figure~\ref{fig:geometry_p}, we derived the following non-linear equations including three known parameters ($s$, $\eta$, $\theta$), and five unknown parameters ($\alpha$, $\beta$, $\xi$, $\phi$, $d$). 

If a local jet structure has a smaller bend; i.e., \rm{$\xi < (\frac{\pi}{2} - \theta$)}, the transformation is:
    \begin{equation}
    \rm{tan\eta=\frac{sin\xi \ sin\phi}{cos\xi \ sin\theta \ + \ sin\xi \ cos\phi \ cos\theta}} \label{eq:1}
    \end{equation}
    \begin{equation}
    \frac{\textit{s}}{\textit{d}}=\rm{cos\beta} \label{eq:2}
    \end{equation}
    \begin{equation}
    \rm{\Big(\frac{tan\beta}{tan\alpha}\Big)^{2}=cos^{2}\eta} \label{eq:3} 
    \end{equation}
    \begin{equation}
    \textit{d} \ \rm{cos\xi \ cos\theta}= \textit{s} \ \rm{cos\eta \ tan\alpha} + \ \textit{d} \ \rm{sin\xi \ cos\phi \ sin\theta} \label{eq:4}
    \end{equation}
    \begin{equation}
    \textit{d}^{2}=\textit{s}^{2} \rm{\Big[\frac{sin\eta}{sin\phi}\Big]^{2}} \ + \ \textit{s}^{2} \rm{\Big[\frac{cos\eta}{cos\alpha}\Big]^{2} \ sin^{2}(\theta+\alpha)} \label{eq:5}
    \end{equation}

If a local jet structure has a larger bend; i.e., \rm{$\xi\geq(\frac{\pi}{2}-\theta$)}, the Equations (\ref{eq:4}) and (\ref{eq:5}) modify as:
    \begin{equation}
    \textit{d} \ \rm{cos\xi \ cos\beta} \ + \ \textit{s} \ \rm{cos\eta \ tan\alpha} = \textit{d} \ \rm{sin\xi \ cos\phi \ sin\theta} \label{eq:6}
    \end{equation}
    \begin{equation}
    \textit{d}^{2} = \textit{s}^{2}\rm{\Big[\frac{sin\eta}{sin\phi}\Big]^{2}} \ + \ \textit{s}^{2}\rm{\Big[\frac{cos\eta}{cos\alpha}\Big]^{2} \ sin^{2}(\theta-\alpha)} \label{eq:7}
    \end{equation}

Our aim is to solve for the angles $\xi$, $\phi$, $\alpha$, $\beta$, and the distance between the knots $d$.
The system is under-determined, and so cannot be solved exactly.
However, as we shall describe, by making use of the angle $\eta$, distance $s$ and angle of LOS $\theta$ as known parameters, we can derive the solution space as well as the relative probability of various bend parameters.\\ 
\indent Please note that \citet{1993ApJ...411...89C} did not give either these equations or a derivation of them, nor did they try to derive more detailed 3-D information about any individual jets. Their aim was to attempt to understand, in a geometrical sense, the misalignment of arcsecond-scale features in blazar jets with the features seen on milliarcsecond scales by VLBI arrays.   
\section{\sl{\textbf{JetCurry}}} \label{sec:jetcurry}
\begin{figure*}[t!]
\centering
\includegraphics[width=1.00\textwidth]{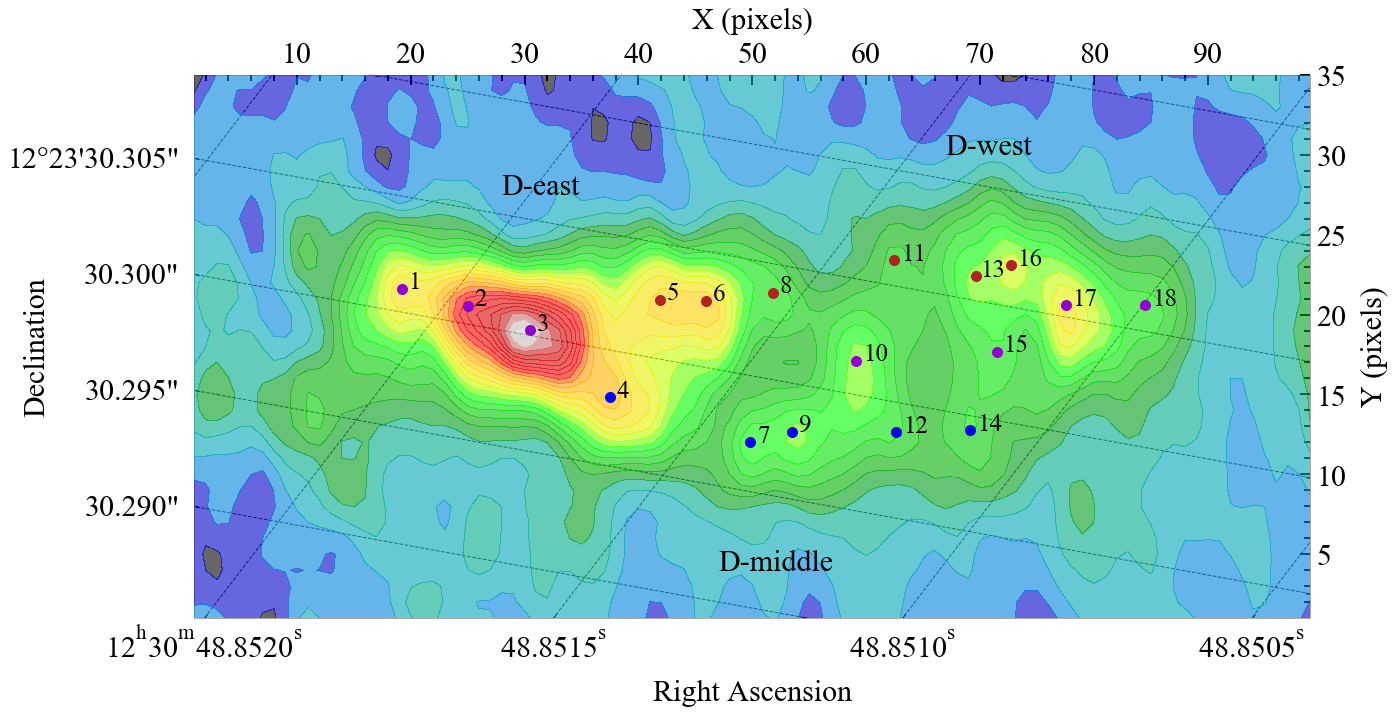}
\caption{The knot D region of the M87 jet. The annotated circular markers correspond to the centroid locations of the identified components, which are acquired from the WISE algorithm at a wavelet scale of 50 mas. The dark red and blue markers belong to the northern and southern streams, respectively. The purple markers are included in both streams. The scale on the image is $0\farcs025/\textrm{pixel}$. The M87 jet's core lies at ($x = -84.23, y = 20.69$) in pixel coordinates, which correspond to R.A. = $12^{\text{h}}30^{\text{m}}49.42^{\text{s}}$ and decl. = $+12^\circ23'28.04''$, respectively. A square root filter is applied to the color scale. The westward direction lies 20.5$^\circ$ below the horizontal. Knot D-east lies between $\approx$15-40 pixels, D-middle lies between $\approx$40-60 pixels, and D-west lies between $\approx$60-80 pixels.}
\label{fig:knotD_outline}
\end{figure*}
{\sl JetCurry} is a Python-based code to visualize the 3-D jet geometry of an AGN jet from its 2-D image. It incorporates the results of the Wavelet-based Image Segmentation and Evaluation \citep[WISE,][]{WISE_Paper1} method as a set of input parameters. These input parameters consist of the centroid locations of the identified components in the 2-D images.
The code uses a nonlinear solving algorithm ({\it emcee}) to solve the non-linear parametrized equations for five unknown parameters: $\alpha$, $\beta$, $\xi$, $\phi$, and $d$ (Equations \ref{eq:1}-\ref{eq:7}). Then, it performs principal component analysis (PCA) and acquires the dominant direction of propagation of the knot D region from the M87's core with respect to our LOS. These tasks are discussed in detail in the following subsections. Our code is available for free as an interactive Jupyter Notebook on Github\footnote[3]{https://github.com/esperlman/JetCurry.}.

\indent We ran {\sl JetCurry} on a radio image of knot D in the M87 jet, chosen as an example of a small region of a relatively complex jet. This is done because the knots have varying and complex flux structures that are separated by considerable distances (see Figure \ref{fig:knotD_outline}). Hence, the jet stream is better visualized by focusing on individual knot regions at a time instead of the entire jet. This restricts the projected jet track from wandering in regions in between knot complexes.

\subsection{Image Decomposition and Component Identification}
{\sl JetCurry} employs the WISE method for multiscale structural decomposition and morphological identification in astronomical images \citep{WISE_Paper1}.
The WISE algorithm implements segmented wavelet decomposition (SWD) to statistically extract significant structural patterns (SSP) at user-specified wavelet scales. It decomposes the image into a set of sub-bands (wavelet scales) and performs a wavelet transform to acquire significant wavelet coefficients against a noise threshold. Then, it uses these coefficients to extract the local maxima coordinates and applies watershed segmentation to retrieve the 2-D boundaries around the corresponding SSP features. 

\indent This methodology successfully identifies complex morphological components, including optically thin and partially overlapping structural features, that are otherwise undetected with standard object-recognition methods \citep[e.g.,][]{BelongieEtAl2002, LobanovEtAl2003, BachEtAl2008}. Figure \ref{fig:knotD_outline} illustrates the centroid locations of the identified SSP features in the knot D region of the M87 jet at a wavelet scale of 50 milliarcseconds (mas). 

\indent The WISE method uses multiscale cross-correlation to detect, classify, and track different SSP features across a series of multi-epoch images. The results of different tests conducted by \citeauthor{WISE_Paper1} (\citeyear{WISE_Paper1}, \citeyear{MertensEtAl2016}) demonstrate the robustness and reliability of the WISE method in identifying and tracking components of the M87 jet that undergo rotation, deformation, and segregation. For the next generation version of {\sl JetCurry}, we will incorporate this capability to analyze the 3-D morphological and kinematic evolution of relativistic jets. 

\subsection{Assumptions, Constraints, and Non-linear Solvers} \label{sec:Emcee and L-LBFGS}
\indent From an observed image in 2-D, we can only measure distances $s$ and angles $\eta$, as shown in Figure~\ref{fig:geometry_p}. We assumed that the LOS viewing angle $\theta$ for the M87 jet is 14$^{\circ}$ with respect to the observer \citep{1999ApJ...520..621B, 2011ApJ...743..119P}. It is important to that
the results acquired from a parsec-scale jet kinematics study of AGNs as part of the 2 cm Very Long Baseline Array (VLBA) survey and Monitoring Of Jets in Active galactic nuclei with VLBA Experiments (MOJAVE) programs suggest that the values of $\theta$ between $70^{\circ}$ and $90^{\circ}$ are very unlikely \citep{ListerEtAl2019}. The best-fit Monte Carlo simulations of the 1.5 JyQC quasar sample indicate that the most jets are viewed at less than $\sim10^{\circ}$. This is also in agreement with the analysis carried out by \citet{VermeulenAndCohen1994}, \citet{ListerAndMarscher1997}, and \citet{Cohen2007}.

The algorithm uses Equations \ref{eq:1}-\ref{eq:7} to solve for the most probable values of the unknown parameters, $\alpha$, $\beta$, $\phi$, $\xi$, and $d$ using a non-linear solver, {\it emcee}, to explore the solution space. {\it Emcee} \citep{2013PASP..125..306F} is a highly efficient, open-source Python package that uses a Goodman $\&$ Weare affine-invarient MCMC Ensemble Sampler, aiming to find a global minimum solution. We used MCMC methods because of the underdetermined nature of the solution space, particularly because of the complex nature of our non-linear trigonometric equations. 
The number of times {\it emcee} is run depends on the user-preference (we are choosing three iterations). The initialization parameters for emcee are 5 dimensions, 1024 walkers, and 50 steps. For our current version of {\sl JetCurry}, we are assuming that the observed jet structures have smaller bends. Therefore, the angles $\alpha$, $\beta$, and $\phi$ are set to be explored from 0 to 1.57 radians, with a step size of 0.5 radians. Consequently, the angle $\xi$ to restricted to [$20^\circ$,\ $\pi/2 - \theta$]. The distance $d$ is set to be floor($s$) to floor($s$)+ 81 parsec (pc). After this, the previous results are set as the initial guess for the next trial until a more defined range of possible solutions is found.\\
\indent Once {\it emcee} has located the global minima/maxima regions, we used a nonlinear solver. To ensure the solutions of the variables are real, we took the logarithm of the equations. We found that the limited memory BFGS algorithm \citep{Broyden, Fletcher, Goldfarb, Shanno} converges to the real solution faster and more accurately in our test cases compared to other nonlinear solvers. The algorithm is in the same class as Quasi-Newton methods and hill climbing with bounds on the solutions. It can be applied to a general non convex function that has continuous second derivatives. In {\sl JetCurry}, the bounds in the nonlinear solver are set to be within 10 steps of the nonlinear solver by default. We present further details of our case study of knot D in \S~\ref{sec:results}.

\subsection{Principal Component Analysis (PCA)}
{\sl JetCurry} applies PCA \citep[see][pg. 289-297, for derivation and algorithm]{Ivezic} and extracts the underlying preferred direction of the knot D region with respect to the LOS. The PCA incorporates the M87 jet's core as the origin and finds a dominant direction of propagation in 3-D. 
This direction is mathematically equivalent to a regression line that minimizes the square of the perpendicular distances from the corresponding Cartesian coordinates \citep[][pg. 292]{Ivezic}. The solid red line in Figure \ref{fig:KnotD_Visualizations} represents the preferred direction of propagation of the knot D region from the M87 jet's core with respect to the LOS ($\theta$ = $14^\circ$). 

\section{Testing{\sl{ J\MakeLowercase{et}C\MakeLowercase{urry}}}} \label{sec:test_JetCurry}
\begin{figure}[t!]
    \centering
        \begin{tabular}{@{}c@{}}
        \includegraphics[width=\linewidth, height = 6.9 cm]{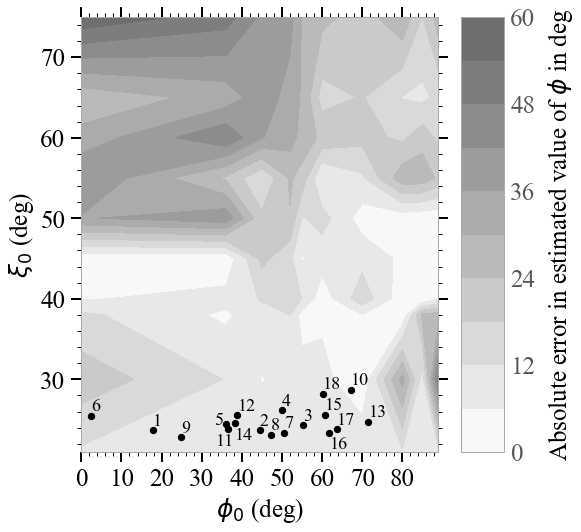} \\
        \includegraphics[width=\linewidth, height = 6.9 cm]{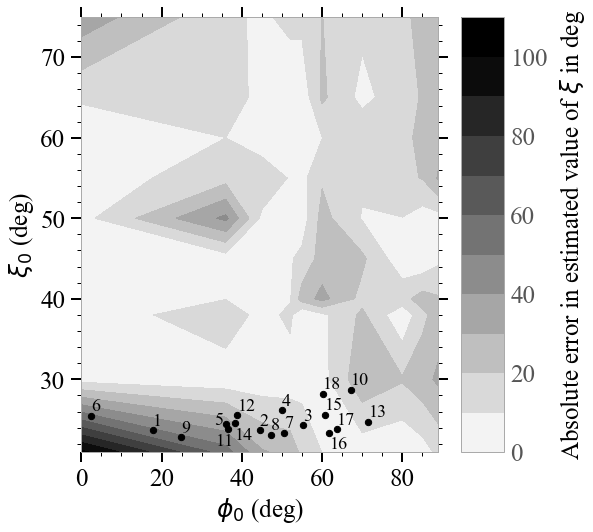} \\
        \includegraphics[width=\linewidth, height = 6.9 cm]{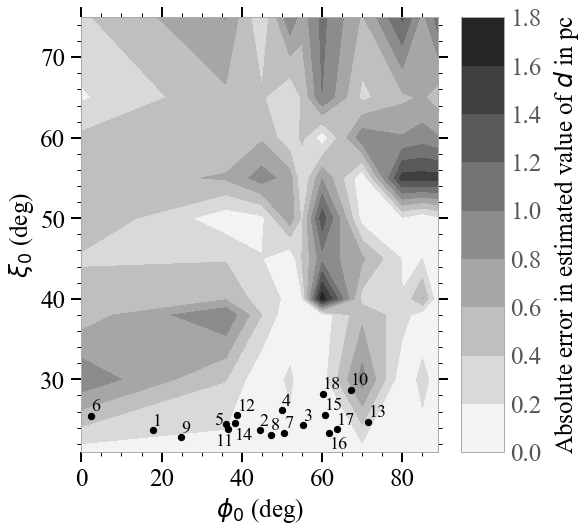}
        \end{tabular}
\caption{Absolute error distribution of estimated values of $\phi$ (top), $\xi$ (middle) and $d$ (bottom) on $(\phi_{0}$, $\xi_{0})$ parameter space. The annotated circular markers represent the acquired values of $(\phi$, $\xi)$ for the identified components in the knot D region of the M87 jet.} 
\label{fig:err_est1}
\end{figure} 
\indent To test the accuracy of {\sl JetCurry}, we simulated 100 bends using different combinations of $\alpha$, $\beta$, $\phi$, $\xi$, and $d$. For our simulated test cases, we assumed that bends greater than 90$^{\circ}$ were most probably unrealistic. We chose $d = 50 \ {\rm pc}$ and $\theta$ = 14$^{\circ}$. Additionally, we restricted the values of $\phi$ and $\xi$ to the range [0,\ $\pi/2$] and [$20^\circ$,\ $\pi/2 - \theta$], respectively.
We then calculated a range of values of ($s$, $\eta$), which are measurable variables in our algorithm. These values of ($s$, $\eta$) were plugged back into our algorithm to test how well we could reproduce the given values of ($\phi$, $\xi$).
 
\indent The resulting values of $\phi$, $\xi$, and $d$ are represented as absolute error distribution plots, which are shown in Figure~\ref{fig:err_est1}. The corresponding absolute errors are given by the color bars. The annotated circular markers are the acquired values of $(\phi$, $\xi)$ for the identified components in the knot D region (see Figure \ref{fig:knotD_outline}). For the cases where the solutions for ($s$, $\eta$) converged successfully, {\sl JetCurry} reproduced the expected bend parameters with the mean absolute errors of $18.16^\circ$, $13.41^\circ$, and 0.49 pc in $\phi$, $\xi$, and $d$, respectively. However, it produced relatively large discrepancies for $0^{\circ} < \phi < 20^{\circ}$ and $ 0^{\circ} < \xi < 20^{\circ}$. This is due to the nature of the equations, and particularly where the expressions in the denominators have singularities. This makes the probability distribution in these regions complex. As a result, MCMC methods do efficiently work backward to find the global minima.

\begin{figure}[t!]
    \begin{center}
    \includegraphics[width=\linewidth]{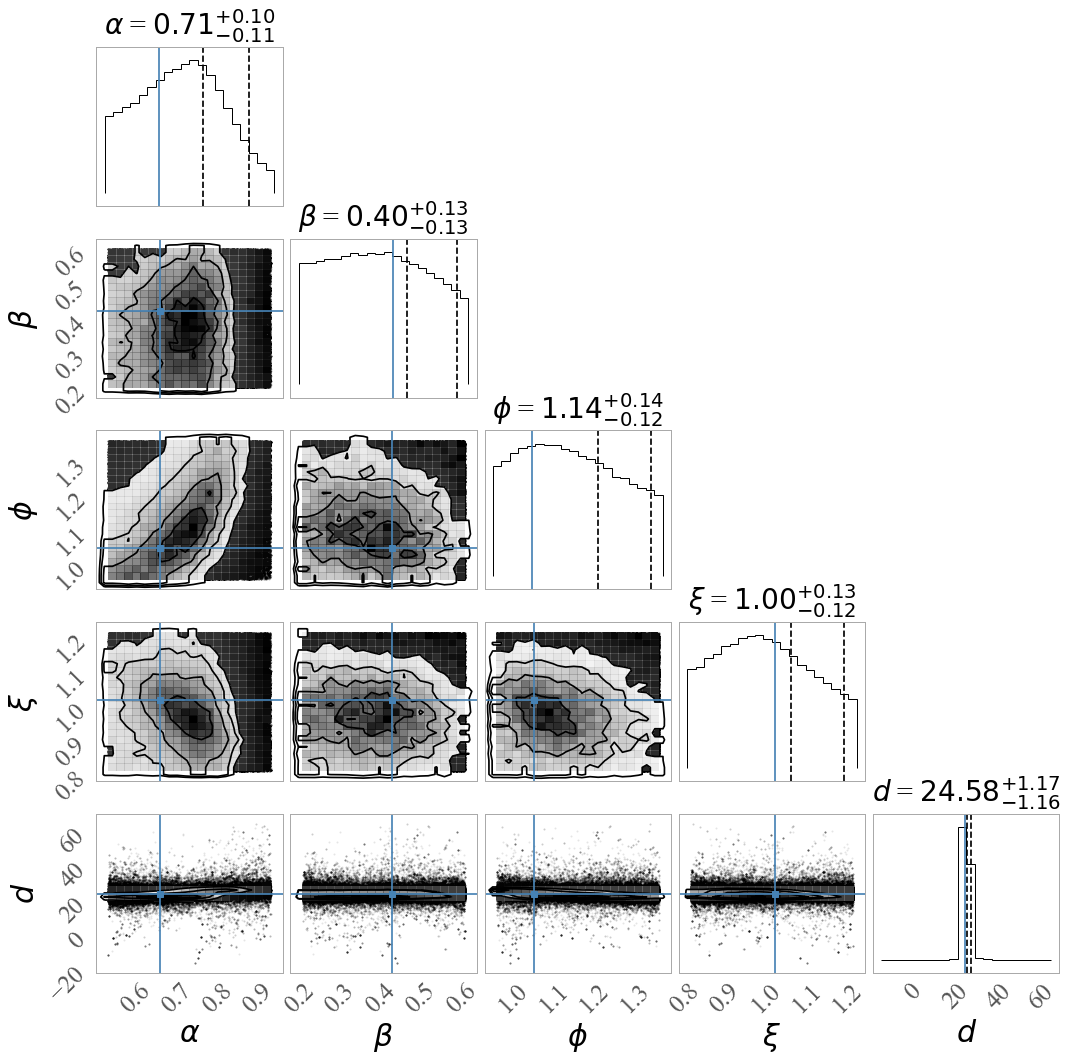}
    \caption{Example of a corner plot showing the marginalized total probability distribution of $\alpha$, $\beta$, $\phi$, $\xi$, and $d$ for a simulated bend with $s = 44.5\ {\rm pc}$, $\eta = 54.98^{\circ}$, and $\theta = 14^{\circ}$. The expected values for ($\alpha$, $\beta$, $\phi$, $\xi$, $d$) are ($40^{\circ}$, $30^{\circ}$, $60^{\circ}$, $60^{\circ}$, $50$ pc), respectively. The solution vector at MAP for $(\alpha$, $\beta$, $\phi$, $\xi$, $d)$ is $(41.22^{\circ}$, $26.49^{\circ}$, $61.98^{\circ}$, $56.06^{\circ}$, $48.86 \ \text{pc})$, respectively. The corresponding absolute errors are ($1.22^{\circ}$, $3.52^{\circ}$, $1.98^{\circ}$, $3.93^{\circ}$, $0.17 \ \text{pc}$). 
    } \label{fig:corner_plot}
    \end{center}
\end{figure}

\indent Figure~\ref{fig:corner_plot} shows a corner plot of the marginalized total probability distribution of $\alpha$, $\beta$, $\phi$, $\xi$, and $d$ for a simulated bend with $s = 44.5 \ {\rm pc}$, $\eta = 54.98^{\circ}$, and $\theta = 14^{\circ}$.
The expected values for ($\alpha$, $\beta$, $\phi$, $\xi$, $d$) are ($40^{\circ}$, $30^{\circ}$, $60^{\circ}$, $60^{\circ}$, $50$ pc), respectively. The solution vector at the location of the maximum a posteriori probability (MAP) for $(\alpha$, $\beta$, $\phi$, $\xi$, $d)$ is $(41.22^{\circ}$, $26.49^{\circ}$, $61.98^{\circ}$, $56.06^{\circ}$, $48.86 \ \text{pc})$, respectively. The corresponding absolute errors are ($1.22^{\circ}$, $3.52^{\circ}$, $1.98^{\circ}$, $3.93^{\circ}$, $0.17 \ \text{pc}$).


\section{Results} \label{sec:results}
\indent We now present a case study of a small region of the M87 jet (i.e., knot D, shown in Figure~\ref{fig:knotD_outline}), which displays a complex morphology.  For this case study, we used the radio image from \citet{2016ApJ...832....3A}. It has a scale of $0\farcs025$/pixel that translates to 2.02 pc/pixel at a distance of 16.7 Mpc. The knot has 3 sub-components, namely, D-east, D-middle and D-west, as well as multiple apparent bends. Knot D-east appears to be at an angle from the northern edge of the jet cross-section to the southern edge over a distance of about 25 pixels ($\sim0\farcs625$). Near its downstream end, it appears to split into northern and southern branches, out of which the southern branch is brighter and has been identified as knot D-middle. 

{\sl JetCurry} uses the centroid coordinates of the identified components (see Figure~\ref{fig:knotD_outline}) to calculate the values of $s$ and $\eta$ with respect to the jet's core $(x_{\rm core}, y_{\rm core})$ and the LOS angle $\theta$ (which we assume is 14$^{\circ}$). It solves the non-linear Equations (\ref{eq:1}) through (\ref{eq:7}) and outputs the range of values  for each unknown parameters, i.e., angles $\alpha$, $\beta$, $\phi$, $\xi$ and the distance $d$. To plot the posterior probability distribution of the values of unknown parameters, we make corner plots for each bend in the knot (similar to Figure~\ref{fig:corner_plot}). Each corner plot is a multi-dimensional representation of projections of the posterior probability distribution of the parameter space \citep{corner}. We interpreted the MAP values as the actual solutions for angles and distances, although this can be subject to irregularities near places where the function is discontinuous or nonlinear (see e.g., \S \ref{sec:VPython}). The gray scale represents the output probability in parameter space, with higher probabilities corresponding to darker colors.
\subsection{Solutions in 3-D Cartesian Space} \label{sec:3-D_soln}
\indent We converted the data obtained from our main algorithm to 3-D Cartesian coordinates $(\textrm{i.e., } x, y, \textrm{and } z)$. These coordinates make use of the most probable values taken from the corner plots obtained for each bend. We used the following transformation equations to calculate the required $(x, y, z)$ coordinates for our 3-D visualizations:
\begin{eqnarray} \label{eq:angle-coord}
x \ &=& \ d \ \rm{cos{\eta}} \ \rm{cos{\beta}} \\
y \ &=& \ d \ \rm{sin{\eta}} \ \rm{cos{\beta}} \\
z \ &=& \ d \ \rm{cos{\eta}} \ \rm{cos{\beta}} \ \rm{tan{\alpha}} 
\end{eqnarray} 
where $\alpha$, $\beta$, $\eta$, and $d$ are the acquired angular bend parameters from {\sl JetCurry}.

\subsection{Visualizations} \label{sec:VPython}
\begin{figure}[t!] 
    \centering
        \begin{tabular}{@{}c@{}}
        \includegraphics[width=\linewidth, height = 5.0 cm]{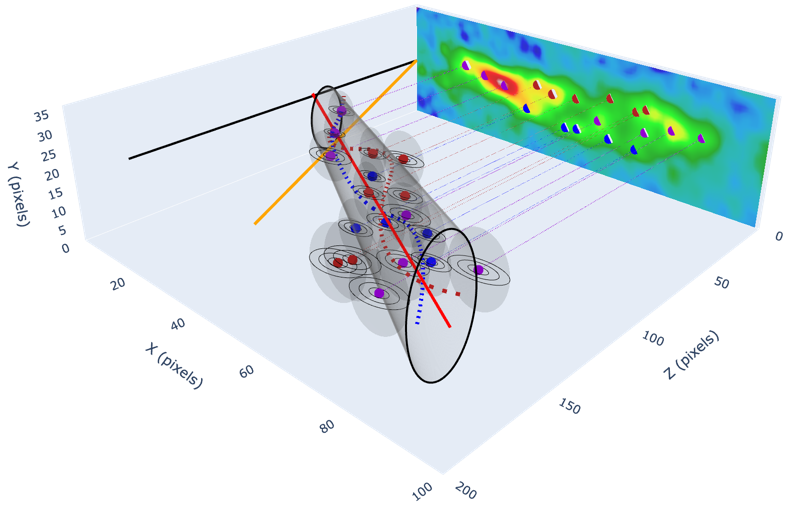}\\
        \includegraphics[width=\linewidth, height = 5.0 cm]{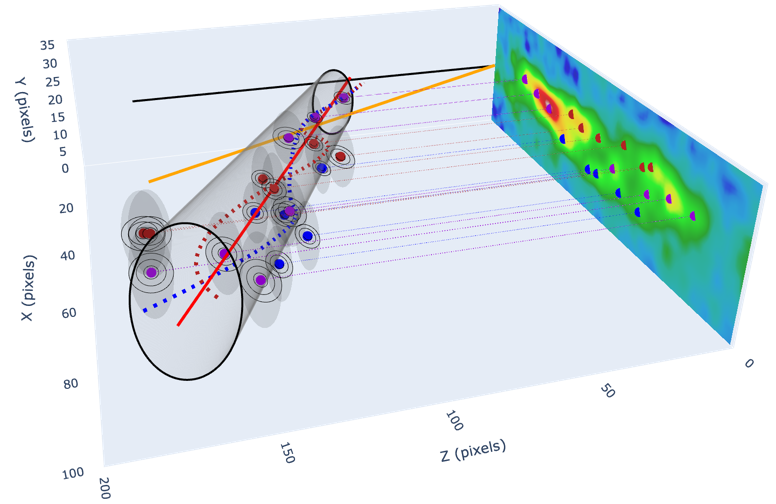}\\
        \includegraphics[width=\linewidth, height = 4.0 cm]{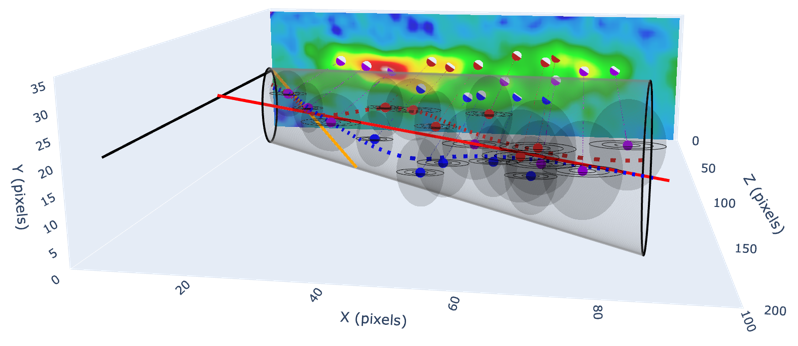}
        \end{tabular}
    \caption{Snapshots of the 3-D visualizations for the knot D region of the M87 jet from three different perspectives. The circular markers represent the most probable solutions of the non-linear parametrized equations. The dark red and blue markers belong to the northern and southern streams, respectively. The purple markers belong to both streams. All circular markers are annotated with $1\sigma$, $2\sigma$, and $3\sigma$ standard deviational ellipsoids, respectively. The dark red and blue dashed curves represent spline fits to the trajectories of the helical paths. The dark red, blue, and purple dotted lines act as a guide to the eye to follow the corresponding projections from 2-D to 3-D. The solid red line represents the preferred direction of knot D from the core ($ x = -84.23$ pixels, $y = 20.69$ pixels) of the M87 jet. The solid orange line corresponds to the LOS with respect to the \textit{z} axis ($\theta$ = $14^\circ$). The 3-D conical structure with a power-law index of $0.96 \pm 0.1$ illustrates the observed shape of the jet in this region \citep{Asada2012}. The scale on the 2-D radio image corresponds to 1 pixel = $0\farcs025$.}
    \label{fig:KnotD_Visualizations}
\end{figure} 

\begin{figure}[t!]
    \begin{center}
    \includegraphics[width=\linewidth]{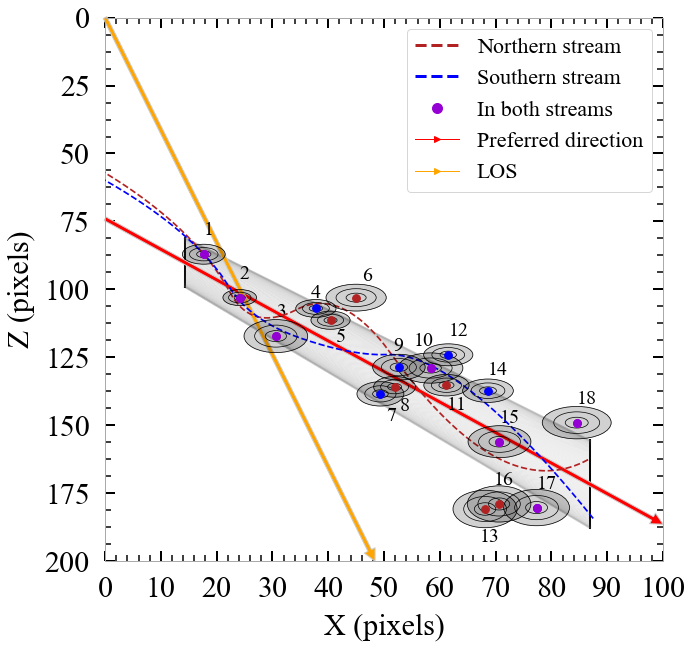}
    \caption{Projection of 3-D visualizations onto the $(x,z)$ plane. The annotated circular markers show the identified components in the knot D region of the M87 jet.}
    \label{KnotD_Radio_2D_Projection_XZPlane_Bins9_WithAllErrors_Annotated}
    \end{center}
\end{figure}

Figure \ref{fig:KnotD_Visualizations} shows 3-D visualizations for the knot D region of the M87 jet. These are snapshots of Plotly's interactive version from three different perspectives. Figure \ref{KnotD_Radio_2D_Projection_XZPlane_Bins9_WithAllErrors_Annotated} illustrates the projection of these 3-D visualizations onto the $(x, z)$ plane.

Each circular marker (dark red, blue, and purple) is a combination of the most probable values of $\phi$, $\xi$ and $d$ for each bend. The dark red and blue markers belong to the northern and southern streams, respectively. However, based on the visual appearance of the knot D region (see Figure \ref{fig:knotD_outline}) and polarimetry data presented in \citet{2016ApJ...832....3A}, we chose a set of points to be in both streams, which are indicated as purple markers. All circular markers are annotated with their corresponding $1\sigma$, $2\sigma$, and $3\sigma$ standard deviational ellipsoids. These ellipsoids signify the spatial dispersion of the acquired bend parameters in the 3-D Cartesian space.

The dark red and blue dashed curves are the spline fits to the northern and southern streams, respectively. We followed the regression analysis in \citet[pg. 193-250]{Gilat}; we found out that these curves are the best fits to the data since they yield the smallest total errors. The solid orange line represents the LOS with respect to the $z$ axis at an viewing angle of $14^\circ$. The solid red line shows the visualized dominant direction of propagation of the knot D region from the M87 jet's core.

Based on the milliarcsecond-arcsecond structure of the M87 jet described in \citet{Asada2012}, we overlaid our visualizations with conical streamlines. Since the knot D region lies downstream of the HST-1 complex, we adopted a 3-D conical structure (shown in gray) with a power-law index of $0.96 \pm 0.1$ \citep{Asada2012}. This helped us further validate the 3-D bend parameters and their corresponding curve fits.

We also made a projection of the acquired 3-D Cartesian coordinates of the bend parameters onto the ($x, y$) plane. To acquire this projection, we used the same equations for $x$ and $y$ as in Equation (\ref{eq:angle-coord}), and assumed $z = 0$. The dark red, blue, and purple dotted lines act as a guide to follow the corresponding projections from 2-D to 3-D. This 3-D interactive plot is available online.


\section{Discussion} \label{sec:discussion}
Based on the most probable values of our bend parameters, {\sl JetCurry} has produced a very interesting 3-D visualization of the knot D region of the M87 jet. It is worth noting that our 3-D visualization is based purely on geometrical considerations.  Thus it is different from physical modeling efforts, which while they start from well known physical principles, have the possible issue that they sometimes do not exactly represent the observed morphologies. A purely geometric visualization can be complementary to such modeling efforts, particularly once the appropriate effects produced by special relativity are included.

While not unique, this 3-D visualization allows us to make several comments regarding the nature of this knot complex. First of all, the acquired 3-D bend parameters are broadly consistent with the observed conical structure of the M87 jet downstream of the HST-1 complex \citep{Asada2012}. 
In fact, it appears likely that the bright regions of knot D lie on at least two filaments (i.e., the northern and southern streams) that are either wrapped around the jet or threading the jet cone, as the two streamlines (Figures 5 and 6) broadly trace the outer edges of the cone.  This supports the idea that  the knot regions represented helical filaments  was first proposed by \citet{1989ApJ...340..698O}. Additionally, the radio and optical polarimetry images presented in \citet{2016ApJ...832....3A} clearly show the apparent helical structure throughout the jet, particularly in the projected magnetic field vectors seen in knots HST-1, D, A and B. Although it is not the only possible way to visualize the jet, it has stood the test of time remarkably well, as discussed by \citet{2016ApJ...832....3A}.

The natural next step in improving {\sl JetCurry} is to incorporate special relativity. If indeed the knot regions are helical filaments wrapped around or threading the jet, a number of special relativistic effects could be observed, such as brightening and apparent acceleration when components' direction of motion crosses the LOS, and dimming and apparent deceleration when components move away from us. This has the ability to break the limitations noted above, as different trajectories would make radically different predictions for the jet's brightness and apparent motion as a function of time.  This could add context to the models that have been proposed  \citep[e.g.,][]{Hardee2011USINGTF} for certain features within knot D as structures downstream of the shock at the eastern end. 

We also plan to use {\sl JetCurry} to visualize the other regions of the M87 jet, especially knots A and B, where our polarimetric images point toward the presence of an intertwined double helix \citep{2016ApJ...832....3A}. This kind of structure is also suggested by previous works. By adding the proper motion constraints \citep[e.g.,][]{Meyer2013}, the observed streamline structures \citep[e.g.,][]{Asada2012}, and the special relativistic considerations (e.g., LOS, Doppler boosting, flux variability, observed superluminal motions, and foreshortening), we can further restrict the ranges of our unknown parameters. These constraints will also help understand the differences in the morphology and hence the emission mechanisms in the inner and outer regions of the M87 jet.


We thank Dr.\,Jeremy Riousset\footnote[4]{Florida Institute of Technology, 150 W. University Blvd., Melbourne, 32901, FL, USA.} for useful discussions on PCA and 3-D curve fits. This research was supported by the National Science Foundation (NSF) grant AST-1716507.

\appendix
\renewcommand{\thefigure}{A-\arabic{figure}}
\setcounter{figure}{0}
\section*{Appendix: Derivation of Non-linear Parametrized Equations} \label{sec:appendix1}
\begin{figure*}[t!]
\centering
\begin{tabular}{@{}c@{}}
    \includegraphics[width=0.9\linewidth]{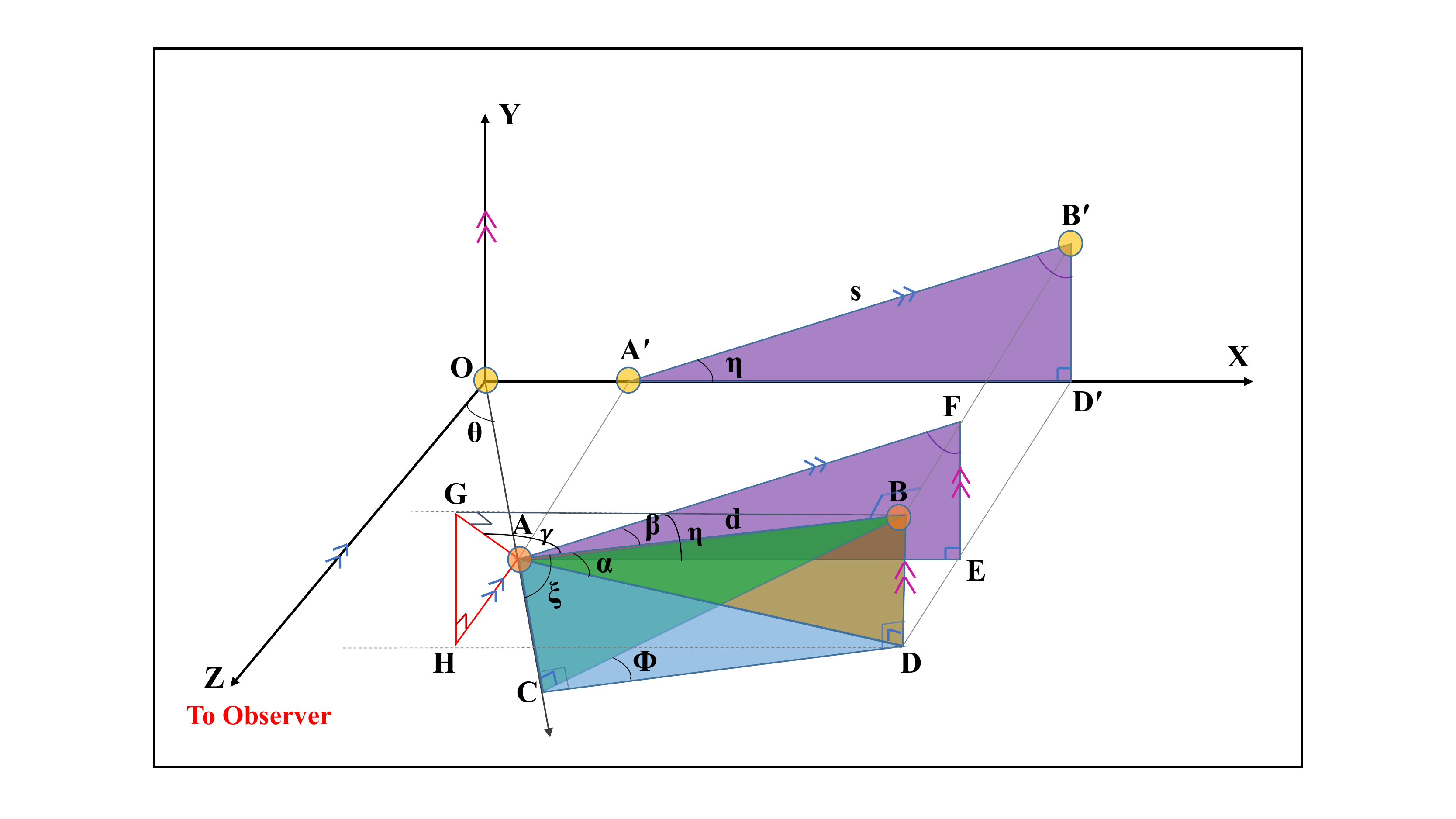} \\
    \includegraphics[width=0.9\linewidth]{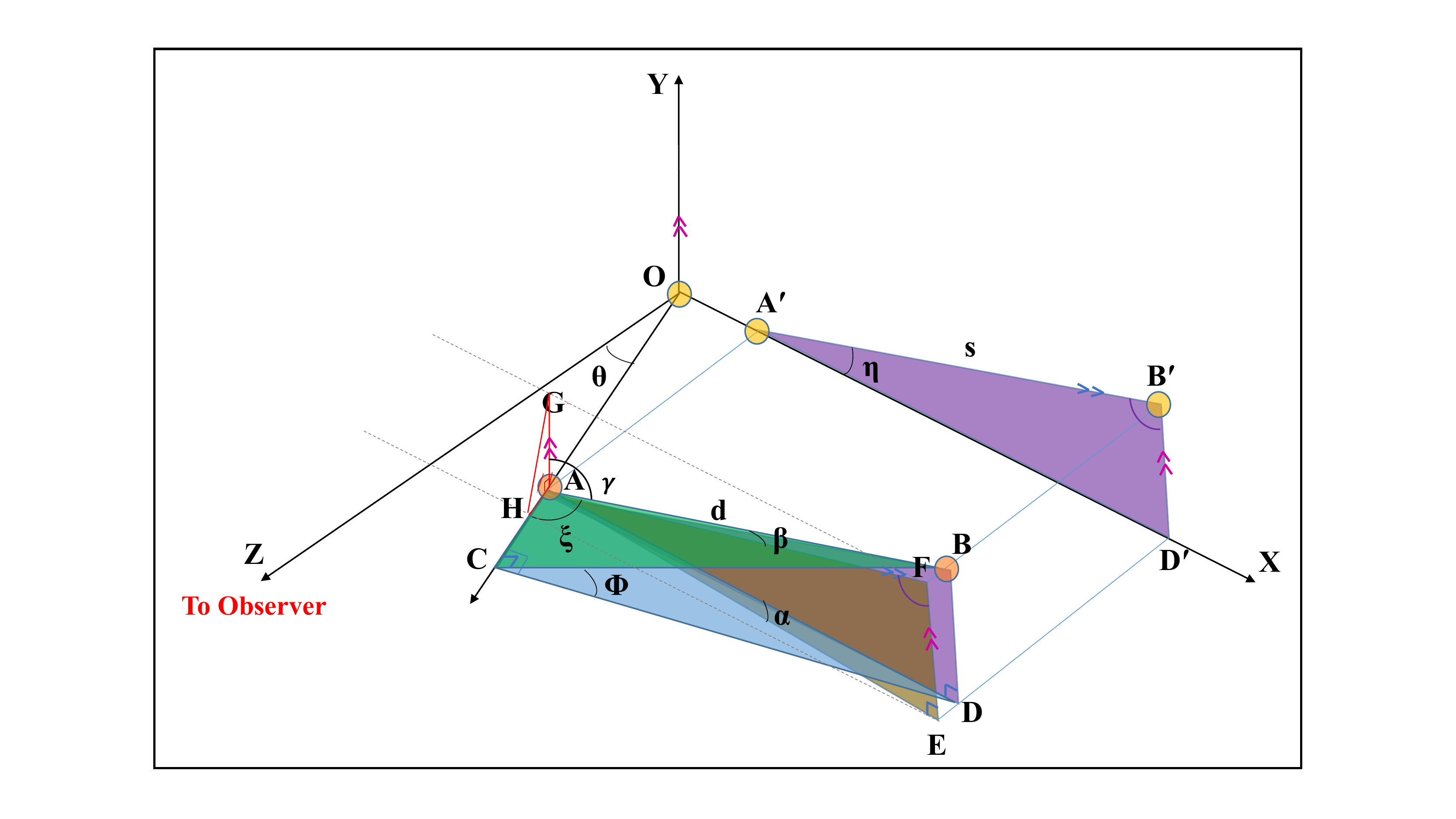}
\end{tabular}
\caption{Visualizing the jet geometry in 3-D from its 2-D projection in the sky frame. Reconstruction of the jet geometry in 3-D for (top) $\xi$ $<$ ($\frac{\pi}{2}-\theta$) and  (bottom) $\xi$ $\geq$ ($\frac{\pi}{2}-\theta$). The jet lies at an angle $\theta$ from the LOS which is assumed along $z$ axis. The projection of the jet in sky frame lies in the $(x, y)$ plane. The points A, B and A$'$, B$'$ represent any two knots in the jet's frame and sky frame, respectively. Point O is an origin and represent the starting point of each iteration of the code. The right angles ($\natural$), equal angles ($\measuredangle$), and parallel lines ($\gg$) are labeled in the figures as appropriate.}
\label{fig:geometry_p}
\end{figure*}
\indent We consider a single bend in the jet at a viewing angle $\theta$ as shown in Figure \ref{fig:geometry_p}. A and B are neighboring knots in the jet's frame, while A$'$ and B$'$ are the projections of knots A and B in the sky frame with the origin at O. The distance between A and B in the jet's frame is $d$ , and the apparent distance between A$'$ and B$'$ is $s$. The projected bend angle relative to the $x$-axis in sky frame is represented as $\eta$. $\angle$BAC is set as $\xi$. 

When $\xi < (\frac{\pi}{2} - \theta$), the equations describing the bend geometry is derived below:

\indent Starting with the equation of \cite{1993ApJ...411...89C} and using the geometry shown in the top plot of Figure \ref{fig:geometry_p}, in $\triangle$A$'$B$'$D$'$, 
\begin{equation}
    \rm{tan \eta = \frac{sin\xi \ sin\phi}{cos\xi \  sin\theta + sin\xi \ cos\phi \ cos\theta}} \label{eq:a1}
\end{equation}
where $\xi$, $\phi$ and $\eta$ are labeled in Figure~\ref{fig:geometry_p}, while $\theta$ is the angle of LOS to the observer. $\triangle$AEF (same as $\triangle$A$'$B$'$D$'$) is the projection of $\triangle$ABD onto the $(x, y)$ plane (i.e., sky frame), and $\triangle$ADE  is the projection of $\triangle$ABF  onto the $(x, z)$ plane. In the right $\triangle$AEF:
\begin{equation}
    AF = s = d \ \rm{cos} \beta \label{eq:a2}
\end{equation}
\indent Because $\triangle$AGH is the projection of $\triangle$ABD onto the $(y, z)$ plane, and $\triangle$AHD is the projection of $\triangle$ABG onto the $(x, z)$ plane. In the right $\triangle$ABG:
\begin{equation}
   BG = HD = A'D'= s \ \rm{cos}\eta \label{eq:a3}
\end{equation}
\begin{equation}
   s \ \rm{cos}\eta = \textit{d} \ \rm{sin}\gamma \label{eq:a4}
\end{equation}
\begin{equation}
    \rm{\gamma= \angle BAG} \label{eq:a5}
\end{equation}
In the right $\triangle$AGH:
\begin{equation}
   \rm{AG^{2} = AH^{2} + GH^{2}} \label{eq:a6}
\end{equation}
\begin{equation}
    AH = DE = s \ \rm{cos}\eta \ \rm{tan} \alpha \label{eq:a7}
\end{equation}
\begin{equation}
    GH = BD = B'D'= s \ \rm{sin}\eta \label{eq:a8}
\end{equation}
\begin{equation}
    d^{2} \ \rm{cos}^{2}\gamma = \textit{s}^{2} \ \rm{sin}^{2}\eta + \textit{s}^{2} \ \rm{cos}^{2}\eta \ \rm{tan}^{2}\alpha \label{eq:a9}
\end{equation}
Equation (\ref{eq:a4}) can be simplified by using Equation (\ref{eq:a2}):
\begin{equation}
   \rm{cos^{2}\gamma = cos^{2}\beta \ sin^{2}\eta + cos^{2}\beta \ cos^{2}\eta \ tan^{2}\alpha} \label{eq:a10}
\end{equation}
Equations (\ref{eq:a3}) and (\ref{eq:a5}) can be combined into:
\begin{equation}
    \rm{\Big(\frac{tan\beta}{tan\alpha}\Big)^{2} = cos^{2}\eta} \label{eq:a11}
\end{equation}
In right triangle CID:
\begin{equation}
    CI = DC \ \rm{sin}\theta = \textit{d} \ \rm{sin}\xi \ \rm{cos}\phi \ \rm{sin}\theta \label{eq:a12}
\end{equation}
and
\begin{center}
\rm{AC cos$\theta$ = AH + CI}
\end{center} therefore;
\begin{equation}
    d \ \rm{cos} \xi \ \rm{cos}\theta = \textit{s} \ \rm{cos}\eta \ \rm{tan}\alpha + \textit{d} \ \rm{sin} \xi \ \rm{cos} \phi \ \rm{sin}\theta \label{eq:a13}
\end{equation}
Also, in right $\triangle$ABC,
\begin{center}
\rm{AB$^{2}$ = AC$^{2}$ + BC$^{2}$}
\end{center} therefore;
\begin{equation}
    d^{2} = \textit{s}^{2} \ \Big[\frac{\rm{sin} \eta}{\rm{sin}\phi}\Big]^{2} + \textit{s}^{2}\Big[\frac{\rm{cos}\eta}{\rm{cos}\alpha}\Big]^{2}\rm{sin}^{2}(\theta+\alpha)  \label{eq:a14}
\end{equation}
\indent From these, we get the following set of five equations, Equations (\ref{eq:a15}) through (\ref{eq:a19}), in five unknowns ($\alpha$, $\beta$, $\phi$, $\xi$, and $d$) along with the measurable parameters ($s$ and $\eta$), and the assumed angle of LOS ($\theta$ = 14$^{\circ}$).
\begin{equation}
    d^{2} = \textit{s}^{2}\Big[\frac{\rm{sin}\eta}{\rm{sin}\phi}\Big]^{2} + \textit{s}^{2}\Big[\frac{\rm{cos}\eta}{\rm{cos} \alpha}\Big]^{2} \rm{sin}^{2}(\theta+\alpha)  \label{eq:a15}
\end{equation}
\begin{equation}
    d \ \rm{cos}\xi \ \rm{cos}\theta = \textit{s} \ \rm{cos}\eta \ \rm{tan}\alpha + \textit{d} \ \rm{sin}\xi \ \rm{cos}\phi \rm{sin}\theta  \label{eq:a16}
\end{equation}
\begin{equation}
    \rm{\Big(\frac{tan\beta}{tan\alpha}\Big)^{2} = cos^{2}\eta}  \label{eq:a17}
\end{equation}
\begin{equation}
    \rm{tan\eta = \frac{sin\xi \ sin\phi}{cos\xi \ sin\theta + sin\xi \ cos\phi \ cos\theta}}  \label{eq:a18}
\end{equation}
\begin{equation}
    s = d \ \rm{cos}\beta \label{eq:a19}
\end{equation}
\indent When the local jet structure has large $\xi$, that is when $\xi \geq (\frac{\pi}{2} - \rm{\theta}$) (the geometry for this scenario is shown in the bottom plot of Figure~\ref{fig:geometry_p}), Equations \ref{eq:a15} and \ref{eq:a16} need to be modified, which become:
\begin{equation}
    d^{2} = \textit{s}^{2}\Big[\frac{\rm{sin}\eta}{\rm{sin}\phi}\Big]^{2} + \textit{s}^{2}\Big[\frac{\rm{cos}\eta}{\rm{cos} \alpha}\Big]^{2} \rm{sin}^{2}(\theta-\alpha)  \label{eq:a20}
\end{equation}
\begin{equation}
    \textit{d} \ \rm{cos}\xi \ \rm{cos}\theta + \textit{s} \ \rm{cos}\eta \ \rm{tan}\alpha = \textit{d} \ \rm{sin}\xi \ \rm{cos}\phi \ \rm{sin}\theta \label{eq:a21}
\end{equation}
\indent The remaining three Equations (\ref{eq:a17}) - (\ref{eq:a19}), stay the same for both the cases. 

\bibliographystyle{cas-model2-names}
\bibliography{The_JetCurry_Code-I}

\end{document}